\documentclass{article}
\usepackage[utf8]{inputenc}
\usepackage{graphicx}
\usepackage{color}
\usepackage{subfigure}
\usepackage[
backend=bibtex,
style=numeric,
citestyle=numeric
]{biblatex}
 
\usepackage{geometry}
 \usepackage{authblk}

\providecommand{\keywords}[1]{\textbf{\textit{Keywords: }} #1}

\title{Future Perspectives of Co-Simulation in the Smart Grid Domain}
\author[1]{C. Steinbrink}
\author[1]{F. Schl\"{o}gl}
\author[1]{D. Babazadeh}
\author[1]{S. Lehnhoff}
\author[2]{S. Rohjans}
\author[1]{A. Narajan}
\affil[1]{OFFIS -- Institute for Information Technology \\
              Oldenburg, Germany\\
              [steinbrink, schloegl, babazadeh, lehnhoff, narajan]@offis.de}
\affil[2]{Hamburg University of Applied Sciences \\
              Hamburg, Germany\\
              sebastian.rohjans@haw-hamburg.de}
\date{June 2018}

\begin{document}

\maketitle

\begin{abstract}
The recent attention towards research and development in cyber-physical energy systems has introduced the necessity of emerging multi-domain co-simulation tools. Different educational, research and industrial efforts have been set to tackle the co-simulation topic from several perspectives. The majority of previous works has addressed the standardization of models and interfaces for data exchange, automation of simulation, as well as improving performance and accuracy of co-simulation setups. Furthermore, the domains of interest so far have involved communication, control, markets and the environment in addition to physical energy systems. However, the current characteristics and state of co-simulation testbeds need to be re-evaluated for future research demands. These demands vary from new domains of interest, such as human and social behavior models, to new applications of co-simulation, such as holistic prognosis and system planning. This paper aims to formulate these research demands that can then be used as a road map and guideline for future development of co-simulation in cyber-physical energy systems.

\end{abstract}

\section{Introduction}
\label{sec:intro}
The energy industry and research is experiencing an ongoing paradigm shift towards cyber-physical energy systems (CPES).
A prominent concept in this development is the so-called smart grid that is often mentioned synonymously with CPES (as done in this work).
The mentioned paradigm shift requires adoption of new concepts for development, testing and validation of these innovative systems and their components. 
The overall validation process spans over multiple stages, starting at mathematical analysis and ending with actual field tests. 
As components in CPES are highly interconnected, test cases quickly become too complex for purely analytical handling.
Therefore, software simulation is an important, intermediate step of the validation process.
CPES include heterogeneous interacting domains like electronics, ICT, automation, politics, economics, energy meteorology, sociology and more.
For several of these domains, simulation tools already exist and are each used for individual studies \cite{pochacker2013,mets2014}.
Integrating these established tools into a combined research environment is the goal of co-simulation.

\begin{figure}[t]
\centering
\includegraphics[width=3.2in]{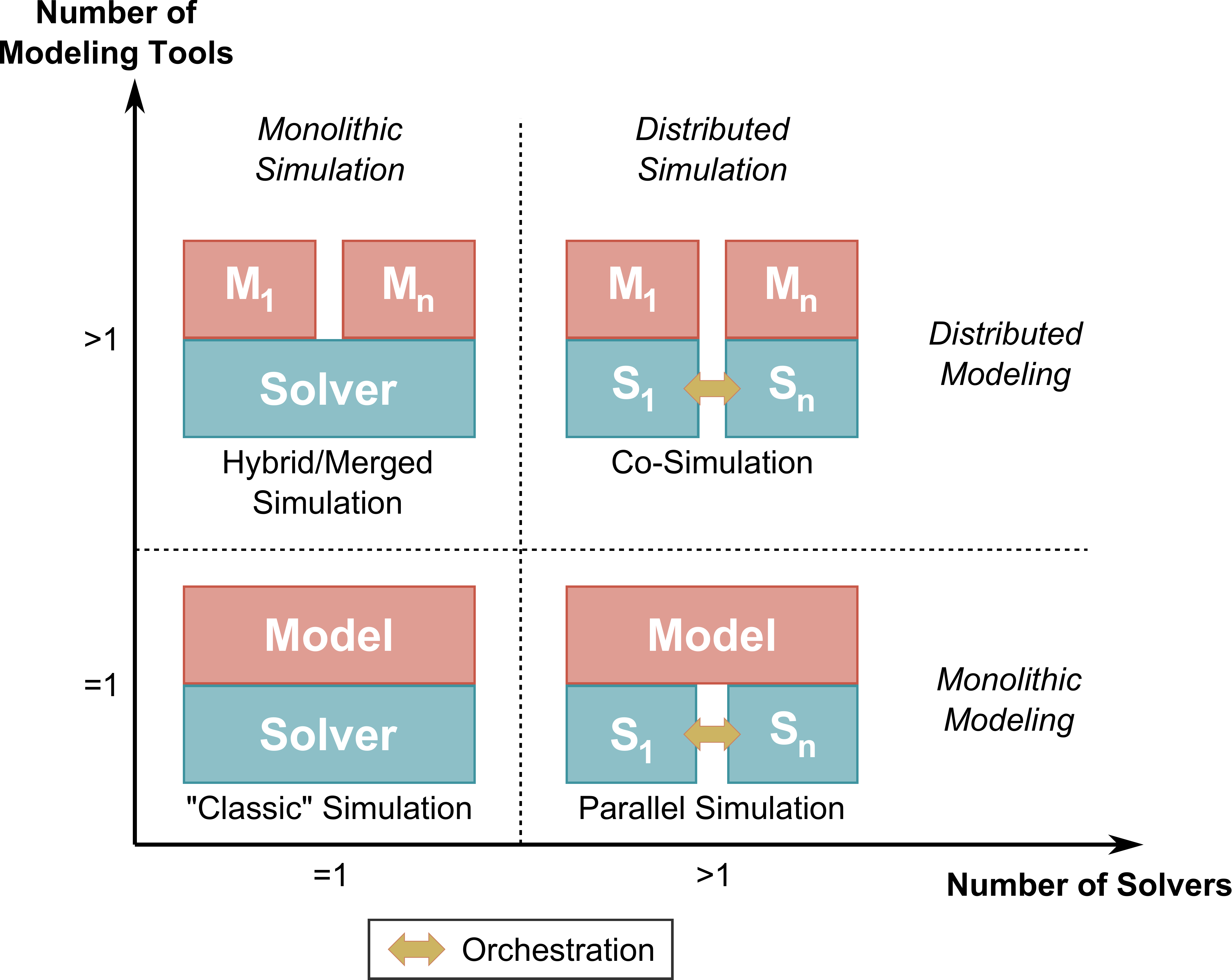}
\caption{Definition of co-simulation (see \cite{schloegl2015}).}
\label{fig_def}
\end{figure}

Co-simulation is defined as the coordinated execution of two or more simulation models that differ in their representation as well as in their runtime environment \cite{schloegl2015}. 
In other words, the models have been developed as well as implemented independently.
In this sense, co-simulation is differentiated from hybrid/merged simulation and parallelized simulation (see Fig.~\ref{fig_def}).
Furthermore, a co-simulation setup may include interactions between hard- and software components. 
If the purpose of a co-simulation scenario is to test hardware, it is typically called \emph{Hardware in the Loop} (HiL) (e.\,g. \cite{de2011}).
Purely software-based co-simulation in the CPES domain is currently mostly focused on analysis of the interaction between power grid and communication network simulators \cite{lin2011,godfrey2010,georg2013,mets2011}.

Co-simulation approaches are subjects of active, ongoing research 
in various domains like the automotive industry, the military, building research, and energy.
Consequently, many generic co-simulation systems already exist, each possessing different features and degrees of usability and popularity. 
Even though co-simulation is gaining more and more momentum, experience shows that various issues are yet to be solved. 
In this paper, the state of the art of co-simulation in CPES is concisely presented and analyzed.
Related review work has been conducted by \cite{Rehtanz_review} and \cite{palensky2017cosimulation}.
However, the paper at hand is more strongly focused on standards and general software tools as well as their applicability.
Furthermore, this paper outlines future development needs in the domain based on real-world requirements as well as experiences gained from different projects.

The paper is structured as follows: In Section~\ref{sec:sota} the state of the art is presented. It is divided into Section~\ref{sec:tools} addressing standards and tools, Section~\ref{sec:req} introducing important non-functional tool requirements and Section~\ref{sec:sg_aspects} analyzing CPES aspects. Section~\ref{sec:fut_dev} derives future development possibilities for co-simulation in CPES. This analysis covers both, advancements of existing approaches as well as new topics that have not been addressed so far. Finally, Section~\ref{sec:conc} concludes the paper.

\section{State of the Art} \label{sec:sota}
Co-simulation research and development is structured into three areas:
functional requirements of tools that widen their applicability, non-functional requirements that improve the general "tool quality", and CPES aspects that are integrated into co-simulation studies. 
All of these areas are connected with one another.
Nevertheless, the following sections address them individually for the sake of structure.

\subsection{Standards and Tools} \label{sec:tools}
The number of co-simulation approaches applied in the power domain has increased so much in the recent years that discussing all of them would require an extensive review article on its own.
This section, instead, aims at providing a conceptual overview of a number of tools and standards with applicability to various types of CPES studies.

One of the most prominent generic co-simulation approaches is the \emph{High-Level Architecture} (HLA) \cite{dahmann1997}, a standard that specifies the interaction of simulation components (called \textit{federates}) managed by a so-called \textit{runtime infrastructure} (RTI).
Several commercial and open source software products support HLA and provide RTI implementations.
HLA simulations are easily distributable by design.
Furthermore, heterogeneous time handling of simulators is permitted by providing a high degree of freedom in the federate interfacing.
HLA is incorporated in a number of tool-specific co-simulation approaches \cite{georg2013, hopkinson2006}.
A rather generic HLA-based platform for smart grid co-simulation is presented by \emph{C2WT-TE}, a web-based framework with integrated community management and a strong focus on simulation as a service \cite{neema2016} (see Sec.~\ref{sec:fut_dev}).

The \emph{Functional Mock-up Interface} standard (FMI) \cite{blochwitz2011} is also widely applied, but follows a different approach than HLA.
FMI specifies a common representation of simulators in the form of \textit{Functional Mock-up Units} (FMUs) that provide a standardized set of interface functions.
A master algorithm that coordinates the FMUs (similar to the RTI) is needed for co-simulation, but is not specified by the standard.
In fact, a combination of HLA and FMI has been suggested by \cite{awais2013}.
However, the more strict FMI specifications for federate design decrease the RTI's potential to handle systems with heterogeneous time representation (e.\,g. event-based and continuous time).
It has been shown by \cite{broman2013} and \cite{broman2015} that such hybrid systems are so far not properly supported by FMI.
Future extensions of the standard may solve this issue.
Nevertheless, FMUs are supported by various generic and tool-specific co-simulation approaches (e.\,g. \cite{bastian2011, rohjans2014}), partly thanks to utility software like the \textit{FMI++} toolbox \cite{widl2013}.

The software \emph{Ptolemy II} \cite{ptolemaeus2014} has been designed primarily for analyzing the interaction of models with heterogeneous structures.
It features a number of scheduler modules (called \textit{directors}) that handle different \textit{domains} (i.\,e. simulation paradigms) and coordinate \textit{actors} (i.\,e. models).
Mapping between the domains allows for hierarchical interaction of different directors.
Although Ptolemy has not been designed specifically for co-simulation, a number of approaches employ it as a framework for energy domain co-simulation \cite{widl2015, Wetter2011}.
However, these setups typically use only one type of director and thus do not utilize Ptolemy's full potential.

A number of more concise co-simulation platforms have been introduced into the energy domain in the recent years.
One notable example is \emph{mosaik} \cite{rohjans2013}, which has been especially designed for high usability.
It features a discretely timed scheduler and a simple interface for simulator integration that supports several programming languages.
A very different type of architecture is realized by the \emph{MECSYCO} \cite{camus2016} approach that provides a multi-agent framework for co-simulation with each agent managing one simulator.
The focus of the approach lies primarily on the interaction of event-based and continuous-time simulation.
Both, mosaik as well as MECSYCO, support FMI \cite{rohjans2014, camus2016b}.

Table~\ref{tab:cosimtools} provides a short categorization of the most important tools mentioned above that are employed as generic co-simulation environments: HLA (as a placeholder for all HLA-based tools), Ptolemy (as a placeholder for all Ptolemy-based tools), mosaik and MECSYCO.
The table indicates that there are several possible criteria for the comparison of such tools that cannot all be mentioned in this work.
Instead, three central criteria have been chosen here as examples: maturity (the time that has already been spent on developing and testing the tools), scope (The extent of the related code and thus the complexity of usage), and architecture (rather centralized with few configuration requirements for new simulators, or distributed with extensive configuration requirements for new simulators).
\begin{table*}[!t]
\renewcommand{\arraystretch}{1.3}
\caption{Comparison of co-simulation environments}
\label{tab:cosimtools}
\centering
\begin{tabular}{c c c}
\hline
Maturity & \shortstack{\textit{Rather recent:} \\ mosaik, MECSYCO} & \shortstack{\textit{Rather established:} \\ HLA, Ptolemy}\\
\vspace*{0.2mm}\\
Scope & \shortstack{\textit{Rather concise:} \\ mosaik, MECSYCO} & \shortstack{\textit{Rather extensive:} \\ HLA, Ptolemy}\\
\vspace*{0.2mm}\\
Architecture & \shortstack{\textit{Rather centralized:} \\ mosaik, Ptolemy} & \shortstack{\textit{Rather distributed:} \\ HLA, MECSYCO}\\
\hline
\end{tabular}
\end{table*}

A special co-simulation requirement is the adherence to real-time constraints used for design and testing of applications varying from wide-area monitoring and control to active distribution grid studies \cite{mikel_platform,Rehtanz_review,davood_book}.
Usability-oriented generic platforms like mosaik are typically rather limited in their real-time capabilities.
There are, however, a number of approaches based on platforms like Ptolemy and HLA that are geared towards hardware integration \cite{Rotger-Griful2016, chemeritskiy2012}.

\subsection{Non-Functional Tool Requirements} \label{sec:req}
Co-simulation approaches are used by domain experts. They want to concentrate on their domain and not on sideline domains like simulation or programming. This makes usability an important requirement for co-simulation tools. Usability includes flexible composition of simulation scenarios and the possibility to reuse existing simulators. This requires standardized interfaces based on common meta information. This meta information is a syntactic and semantic description of the data a simulator can provide and receive. Such a description also helps to avoid mistakes when connecting individual simulators \cite{schutte2013simulation}. Scenario meta data can be considered as a set of rules for component composition in a co-simulated system. Incorrect connection of simulators may be avoided this way.
Even more so, appropriate formalisms enable the automatic generation of scenarios \cite{schutte2011domain}.                                

Next to usability, performance is a constant issue for simulation tools.
Smart grid simulations usually deal with large and complex systems that require a lot of resources in terms of hardware and calculation time. Co-simulation comes with additional costs for data exchange and synchronization of the individual simulators. On the other hand, co-simulation also offers different approaches for performance improvement, most notably the possibility to run the individual simulators in parallel, even distributed over several computers. Another approach is the replacement of complex and slow simulators with faster surrogate models \cite{simpson2001metamodels}. Such surrogate models are usually less accurate and/or only work in a certain range of system states.
This, however, is acceptable if the focus of the analysis lies on other parts of the simulated system.

Lacking accuracy in surrogate models or any other type of simulators entails the need for an uncertainty quantification (UQ) process.
As the name suggest, this process allows to assess the uncertainty/accuracy of simulation results when using them to derive real-world predictions.
The large, mathematical UQ community has already established a wide variety of methods and tools \cite{uqbook}.
However, application of these techniques in the co-simulation domain has so far not received much attention.
One of the first systematic UQ approaches to smart grid co-simulation is provided by the \emph{MoReSQUE} module that has been suggested to extend mosaik \cite{Steinbrink2016b}.

\subsection{Smart Grid Aspects} \label{sec:sg_aspects}
CPES co-simulation has so far played a significant role in the analysis of the integration of ICT systems and new services into conventional power systems.
For example, co-simulation testbeds have been used in literature to analyze and quantify the performance of wide-area monitoring as well as control and protection applications in real power system scenarios and the impact of the supporting ICT \cite{powertech2013,wide_area1}.
From a market perspective, co-simulation has opened the door to more sophisticated and futuristic studies such as heat trading in a power grid considering the thermo-hydraulic properties of the grid \cite{heat_trading}.

\section{Possible Future Developments} \label{sec:fut_dev}
\begin{figure*}[t]
\centering
\includegraphics[width=0.7\textwidth]{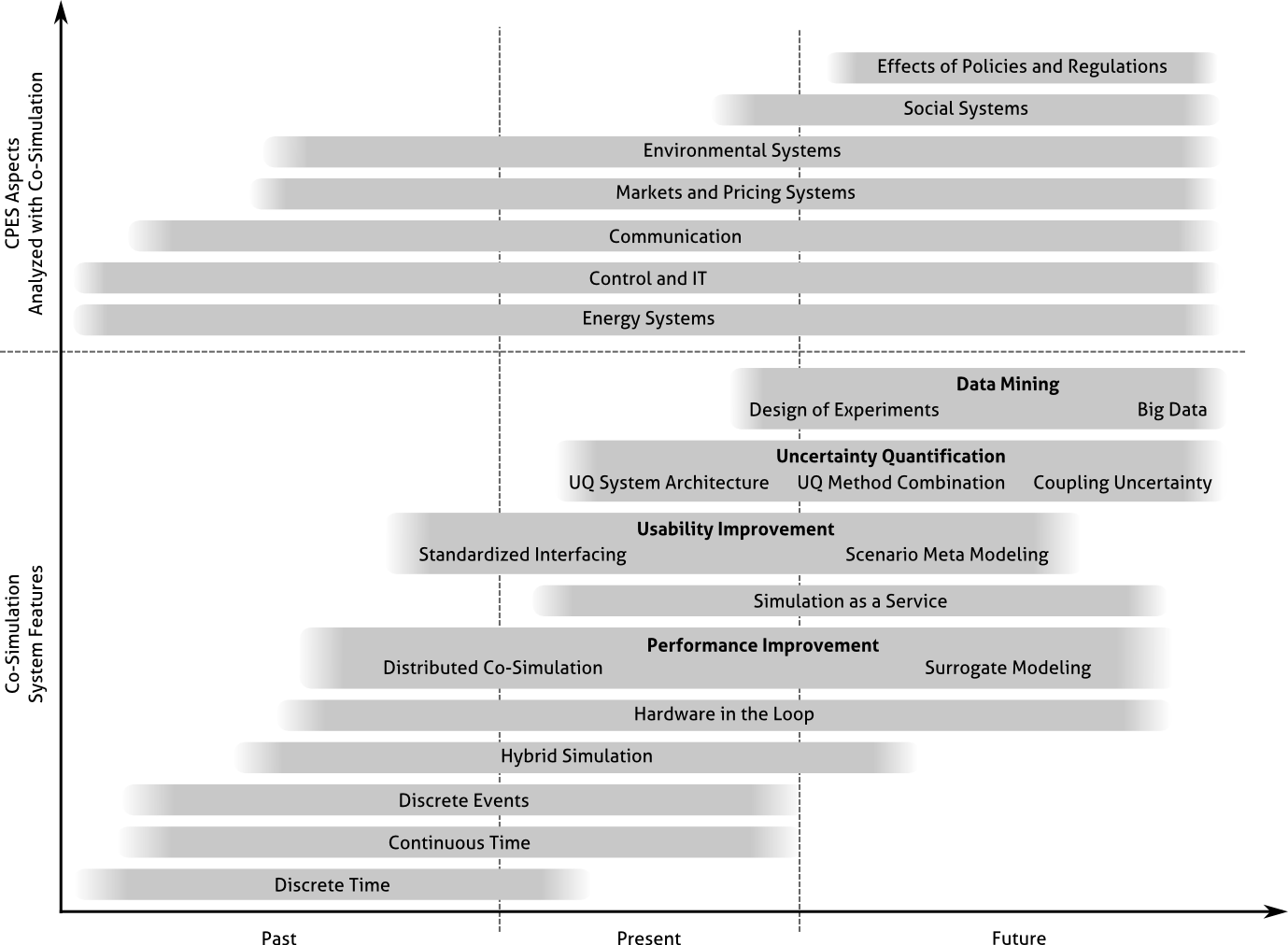}
\caption{Overview of co-simulation topics in CPES.}
\label{fig_sim}
\end{figure*}

Co-simulation is a rather recent subject in CPES research.
Despite the effort that has already been invested in the topic, many possible applications and improvements are still on the horizon.
This section attempts to provide a broad overview of the potential future development of the domain.
Figure~\ref{fig_sim} depicts a rough temporal projection of these developments in terms of improvements of co-simulation technology as well as fields of application for CPES co-simulation.

The strongly interdisciplinary character of CPES is so far not fully reflected in the contemporary co-simulation studies.
Typically, co-simulation approaches focus only on a subset of the relevant smart grid domains.
As an example, studies may deal with the interaction of power grids and communication systems (see \cite{mets2011}), or energy consumption as influenced by market dynamics (e.\,g. \cite{awais2013}).
In real-world systems, however, the number of influencing domains cannot simply be reduced.
Therefore, prospective simulation systems should be able to analyze the interaction of an increasing number of systems in order to make meaningful predictions about smart grid dynamics, including consideration of social, economic and environmental effects.
As one example, advanced topics such as cyber security and transactive energy system should be kept in mind when designing futuristic co-simulation platforms \cite{TES}.
Furthermore, transactive energy systems are intensively influenced by market-based response and human behavior. 
Especially human behavior is a critical factor for power system operation.
Examples span over the interaction of operating crews in control rooms, human error in setting up components and tools at substation, the behavior of cyber attackers, or consumer behavior for demand-respond management \cite{human}. 

More and more complex co-simulation studies will lead to interaction of various simulation paradigms, especially discrete-time, continuous-time and discrete-event simulation.
Thus, co-simulation platforms should provide versatile scheduling possibilities and also support hybrid simulation.
One of the major challenges in modern co-simulation is to combine such capabilities with high usability.
Tools like mosaik are strongly tilted towards usability with a slim simulator interface designed for a single scheduler type.
FMI, while being slightly more complex and powerful, is still not properly adapted to hybrid simulation.
HLA and Ptolemy, on the other hand, are capable of handling various simulation paradigms but provide less well specified interfaces with higher implementation effort.
A possibility to bridge the gap between usability and versatility is to provide various scheduler modules with associated interface options of different complexity.
Model providers may then choose to implement the interface and employ the scheduler that suits their simulators.
Hierarchical mapping between schedulers can be employed to realize hybrid simulation, as it is already supported by Ptolemy.
The different interface options should be compliant with FMI, possibly in an extended version, for the sake of standardization.

Next to scheduler modularity, improvement of co-simulation platforms also brings up the issue of architectural choices.
Systems with centralized scheduling like Ptolemy or mosaik provide the advantage of higher reusability in the creation of various co-simulation studies.
Decentralized setups like MECSYCO, on the other hand, are typically more easily distributed, allowing for easier parallelization and thus performance increase. 
A middle ground is possible by attuning scheduler modules to distribution-friendly simulator management systems like HLA's RTI.

As mentioned before, HiL studies entail some special requirements for co-simulation setups.
Currently, HiL tests are usually carried out either using real-time co-simulation for tests with hard real-time requirements (e.\,g. for power electronics testing), or using offline co-simulation for soft real-time requirements (e.\,g. for EMS applications) \cite{Rehtanz_review}. 
However, future co-simulation platforms/architectures should be adopted to deal with more sophisticated hardware requirements. 
For example, current FPGA-based real-time co-simulation has limitations in terms of dealing with the frequent changes of switching topology in power electronic-based HiL tests.

Since the CPES co-simulation of the future will entail more and more large-scale and complex scenarios and special timing requirements, performance becomes even more important.
One possibility to improve  performance is to use faster models. 
By using simulation models with different granularity for the same component it is possible do adapt to requirements on precision and calculation speed, using more sophisticated models if high precision is requested, or simpler models if calculation speed is most relevant. 
In a further step this can be automated via performance profiling and performance classification of simulators. 
Creation of faster surrogate models is typically done via machine learning approaches, e.\,g. artificial neuronal networks \cite{simpson2001metamodels}, that are trained based on input and output data.
Even more so, such approaches can be used to create combined representations of simulator sets in co-simulation setups.
Complete subsystems of CPES may be modeled within one (surrogate) simulator instead of being composed of several.
This may be applied to co-simulation studies that are strongly focused on certain parts of the smart grid.
The peripheral parts may be represented by a small number of more coarse models.
It is particularly useful in large-scale simulation setups.

As mentioned before, users of co-simulation tools are usually no simulation experts or even advanced programmers. 
Hence, usability is an important field for future development. 
Standardized meta information that describes the data simulators can exchange help to avoid mistakes when interconnecting the simulators. 
An exact identification of the data helps to avoid wrong connections, e.\,g. the connection of a sensor's power source to a high voltage line.
To facilitate the generation of scenarios, i.\,e. the combination of individual simulation models to an overall co-simulation, a scenario meta description is useful. 
As described in Section~\ref{sec:sota}, first approaches of simulator and scenario information models already exist.
However, these approaches do not yet account for the full complexity and future extensibility of smart grid systems.
Appropriate new standards may be used for co-simulation automation and validation.

Easily usable simulation environments could even make co-simulation accessible to users in decision maker positions.
These users are typically not trained programmers or even researchers.
Therefore, co-simulation tools for decision making should be equipped with features like the already mentioned simulation automation and comprehensive graphical user interfaces (GUI).
Aside from usability, a holistic smart grid outlook is especially relevant for decision makers.
Such stakeholders are typically assigned with planning or regulatory tasks that may influence the system in question on a large scale so that it is important not to overlook possible side effects of their decisions.
Given such an appropriate setup and corresponding simulators, co-simulation may lead the path to enhanced CPES planning.
Planning concepts that so far have been handled mainly theoretically can then be pursued much more rigorously via simulation-based methods.
An example is the so-called \textit{backcasting} that has already been discussed in the context of smart grid planning \cite{robinson1982, nathwani2010}.
It is used to propose a desired future system state and then analyze possible steps to reach it based on the current state.
Energy backcasting that is supported by co-simulation studies requires appropriate co-simulation environments and simulators, as well as an overarching concept that connects the theoretical approach with the calculations.

Increasing application of co-simulation in an industrial or regulatory context entails the need for more thorough UQ.
The already mentioned MoReSQUE concept suggests a modular approach for UQ in smart grid co-simulation.
This would allow for the coupling of different UQ methods within the same study with each simulator being handled by the method that suits it the most.
However, such an UQ method coupling has so far not been thoroughly conducted so that potential issues and merits have to be analyzed first.
Furthermore, co-simulation is afflicted by the uncertainty source of simulator coupling.
Approaches like \cite{arnold2013} provide analysis of numerical coupling errors in FMI-based co-simulation.
Based on such groundwork, systematic description and handling of coupling uncertainty can be developed that is then to be combined with UQ of model and data uncertainties.

In some use cases it is useful if a co-simulation system is able to ``replay'' certain events. An example is a simulation environment that runs in parallel to a real system, e.\,g. as part of an assistance system that helps the operator in controlling the power system. 
In such a case, it may be required to replay a critical situation in order to analyze it and to take appropriate measures for the future. This is mainly a matter of data management. As this example is about unforeseen events, as much data as possible has to be stored, which means that huge amounts of data have to be handled.
In other words, future co-simulation development will also have implications on data science and data management needs.

\section{Conclusion} \label{sec:conc}

The presented review work gives a very concise overview of the CPES co-simulation domains. 
The major contribution of this paper, however, lies in the outlining of possible developments in the future of smart grid co-simulation.
These perspectives have been derived from real-world and research project requirements.

One major challenge that has been identified is the reconciliation of hybrid coupling capabilities with high framework usability that boost collaboration as the core merit of co-simulation.
This demand is valid for the work of researchers (standardized interfacing, scenario validation) as well as decision makers (GUIs, analysis tools).

Next to that, additional research is needed for features that improve result quality and quantity of simulation studies, namely performance enhancement (e.\,g. through distribution or surrogate modeling) and UQ.
Once large sets of high-quality simulation studies can be conducted quickly, CPES co-simulation becomes more interesting for the field of big data management and data mining.

With more capable co-simulation environments, more and more subordinate CPES domains may be included into common research studies.
This will lead the way to increasingly holistic simulation and thus simulation-based system planning.

All in all, the field of CPES co-simulation is too diverse to be covered in its entirety by a single review article.
In particular, a comprehensive overview of available tools would require a paper on its own.
One important topic that has been omitted here for the sake of scope is educational needs in the CPES and co-simulation domain.
Obviously, the associated challenges are strongly related to the issue of usability in co-simulation (see e.\,g. \cite{strasser2014edu}).

\section*{Acknowledgements}
This work is supported by the European Community’s Horizon 2020 Program (H2020/2014-2020) under project ``ERIGrid'' (Grant Agreement No. 654113).

\printbibliography

\end{document}